\documentclass[11pt, oneside]{article}   	
\usepackage{geometry}                		
\usepackage[parfill]{parskip}    		
\usepackage{graphicx}				
\usepackage{amssymb}

\title{Disambiguation of Patent Inventors and Assignees 
Using High-Resolution Geolocation Data}
\author{Greg Morrison\thanks{IMT Institute for Advanced Studies, Lucca, Italy} \and
Massimo Riccaboni\thanks{IMT Institute for Advanced Studies, Lucca, Italy and 
Department of Managerial Economics, Strategy and Innovation, K.U. Leuven, Leuven, Belgium.} \and
	Fabio Pammolli\thanks{IMT Institute for Advanced Studies, Lucca, Italy}}

\def\Pv{{\mathbf{P}}}
\def\nv{{\mathbf{n}}}

\begin{document}
\maketitle
\begin{abstract}
\noindent Patent data represent a significant source of information on innovation and the evolution of technology through networks of citations, co-invention and co-assignment of new patents.  A major obstacle to extracting useful information from this data is the problem of name disambiguation:  linking alternate spellings of individuals or institutions to a single identifier to uniquely determine the parties involved in the creation of a technology.  In this paper, we describe a new algorithm that uses high-resolution geolocation to disambiguate both inventor and assignees on more than 3.6 million patents found in the European Patent Office (EPO), under the Patent Cooperation treaty (PCT), and in the US Patent and Trademark Office (USPTO). We show that our algorithm has both high precision and recall in comparison to a manual disambiguation of EPO assignee names in Boston and Paris, and show it performs well for a benchmark of USPTO inventor names that can be linked to a high-resolution address (but poorly for inventors that never provided a high quality address).   The most significant benefit of this work is the high quality assignee disambiguation with worldwide coverage coupled with an inventor disambiguation that is competitive with other state of the art approaches.  To our knowledge this is the broadest and most accurate simultaneous disambiguation and cross-linking of the inventor and assignee names for a significant fraction of patents in these three major patent collections.   
\end{abstract}

\section{Introduction}

In many contexts, technological progress and innovation is essential to national or regional economic growth and output.  One way of measuring innovation is the production of patents, which represent a technological advancement produced by individuals (generally, these are inventors on the patents) working at research institutions (generally, these are the assignees on the patent).  The analysis of patent databases has provided techniques for evaluating information spillovers \cite{spillovers,miguelez}, inventor mobility between regions \cite{marx2007noncompetes,mobility}, interregional and international collaborations \cite{bordercollab,mansci,riccaboni,morescalchi,morrisonjcn}, and the emergence of new technologies \cite{youn2015invention}. Among many others, these studies have provided an in-depth picture of the dynamics of regional and institutional talents and quantify the success of various inter-institutional and inter-regional collaboration, of great use to policy makers.

A major problem in the use of patent data (or any bibliometric database, such as for scholarly publications \cite{authority}) is the disambiguation of authors or institutions.  There are a wide range of alternate spellings of a person's or institution's name, where, for example, ``The National Institutes of Health'' and ``NIH'' may refer to the same institution.  Typos and misspellings of names are also common in bibliographic data (e.g. ``National {{Institute}} of Health,'' missing an `s' in the second word, is assignee on 24 patents worldwide).  The goal of disambiguation is to link all of these alternate spellings of institutional or individual names {\em{without}} incorrectly linking similar names referring to distinct entities.  This is a difficult task, as there are millions of names to disambiguate (making pairwise comparisons of the full dataset computationally expensive) and an evaluation of how likely  two names on patents are to be the same entity is not known a priori and often relies on machine learning techniques \cite{authority,inventorDisambig,assigneeDisambig2,inventorDisambig1}.  

In this paper, we describe a straightforward but accurate approach to the disambiguation problem using high precision geolocation of assignee and inventor addresses.  Two inventors (or two assignees) that provide exactly the same high-resolution address and {\em{also}} have `similar' names are very likely to refer to the same entity.  Knowing that two entities have exactly the same address allows a great deal of flexibility in name matching, and we design two simple string matching approaches to link similar names that share a high-resolution geolocation.   Inventors and assignees that have addresses with low resolution, which are extremely common in the USPTO data ($\sim 95\%$ of inventor addresses have no street provided), are linked to exact name matches nearby, greatly increasing the coverage of the disambiguation.  We show that this approach provides a complete, high-resolution disambiguation of inventors and assignees on 3.6 million patents, and that the precision and recall of the resulting disambiguation is superior to or competitive with other well known disambiguation methods on that subset of the patent data.  

The primary advantages of this methodology over existing disambiguation approaches are threefold.  First, the assignee disambiguation appears robust and of a broader scope than is found in other methodologies in the literature\cite{assigneeDisambig1,assigneeDisambig2}.  The assignment of unique institutional identifiers to these assignee names will allow a better understanding of inter-institutional mobility of inventors and institutional collaborations, a significant benefit provided by this dataset.   Second, the fact that the inventor disambiguation works well in comparison to more refined inventor disambiguation techniques\cite{inventorDisambig} over a significant fraction of patents worldwide suggests that our dataset can be useful in understanding inventor collaborations or mobility.    Finally, the use of both US- and European-centric patent offices ensures that we have a global focus in our data, with a good chance of linking assignees and inventors found in multiple offices at similar addresses.

\section{Overview}

\begin{figure}[htbp]
\begin{center}
\includegraphics[width=.42\textwidth]{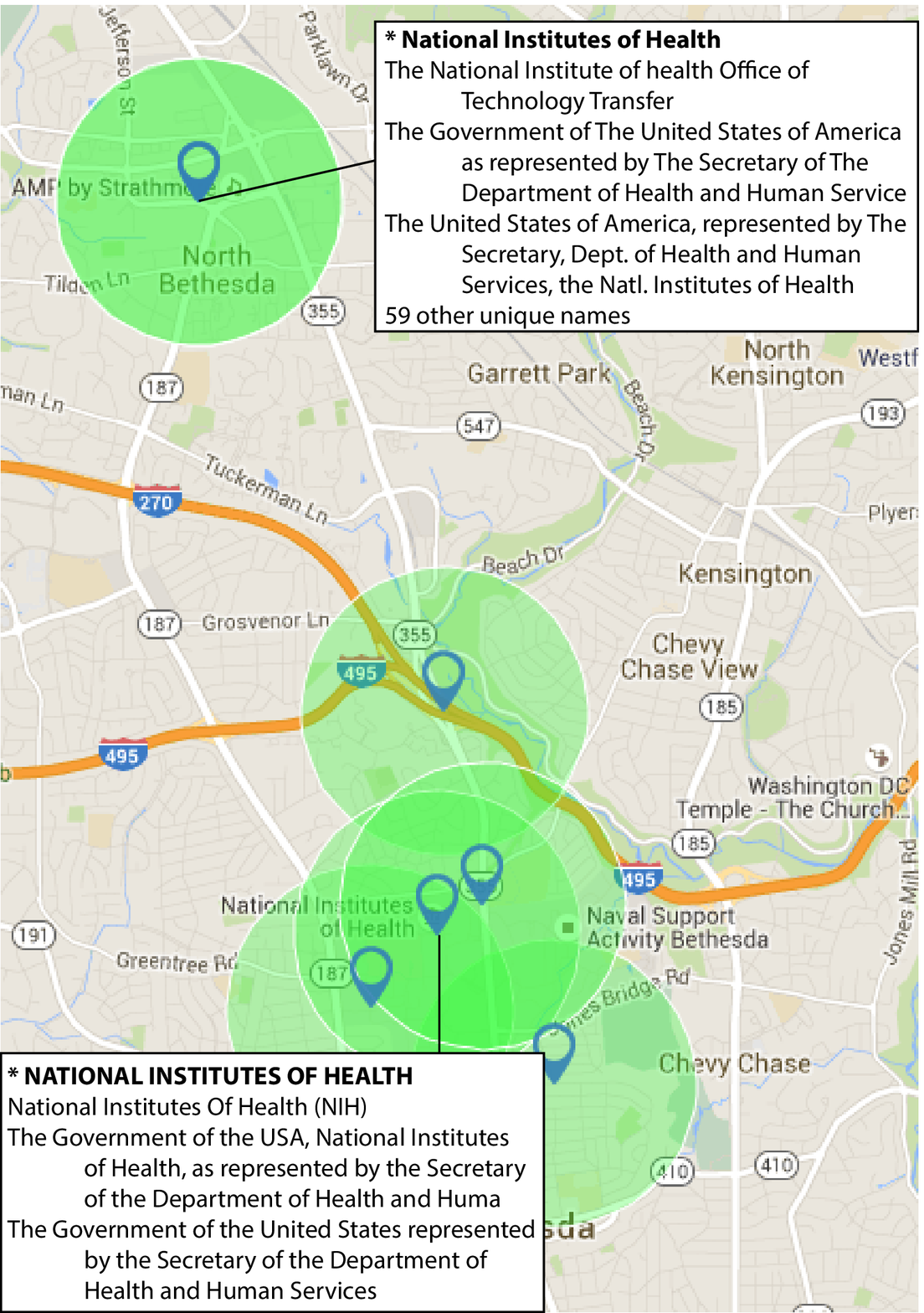}\quad\includegraphics[width=.55\textwidth]{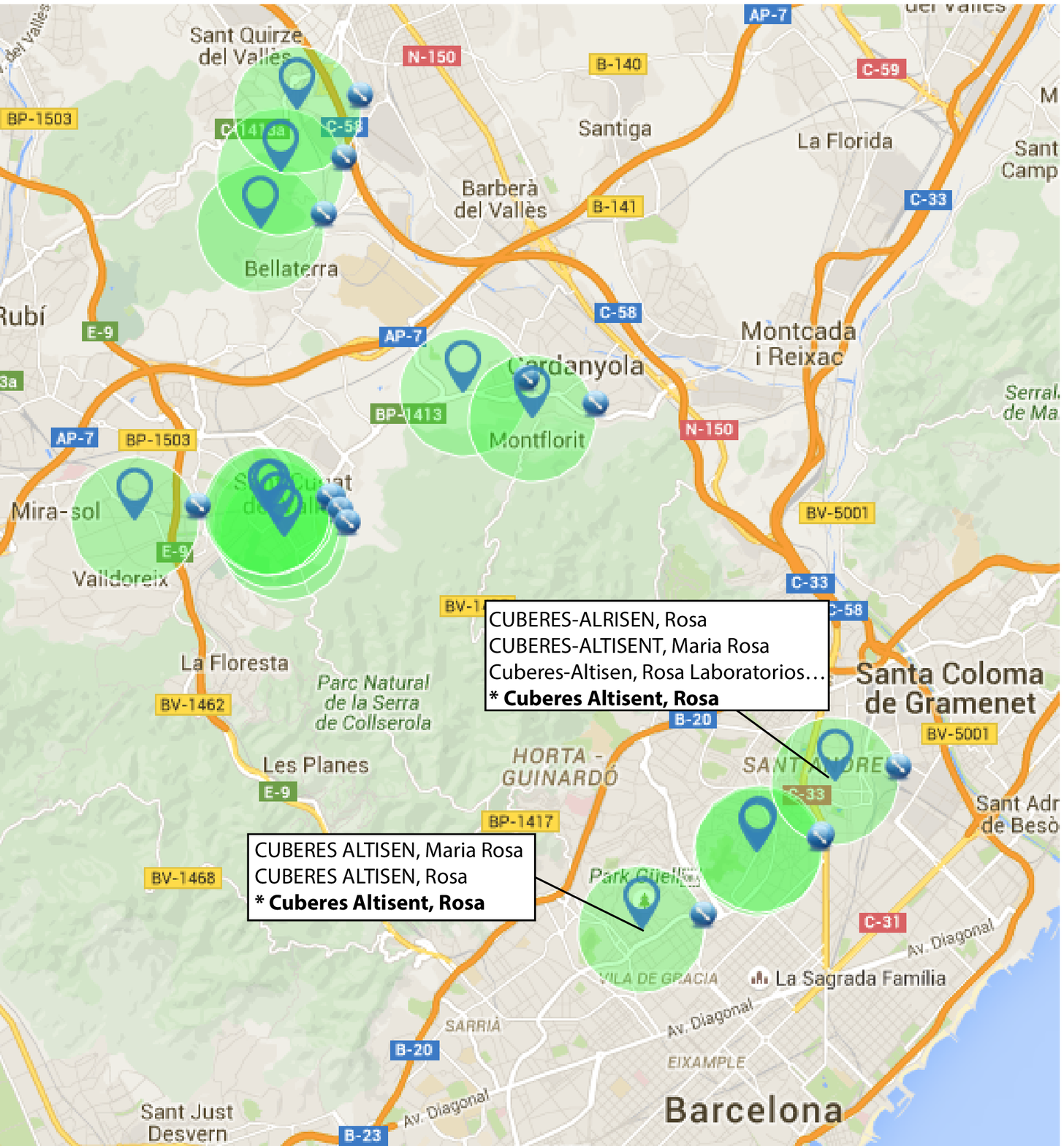}
\caption{Geolocations of assignee addresses (left) or inventor addresses (right) for two examples:  the NIH in Bethesda Maryland, and an inventor in the Barcelona area with many names and addresses.  In both cases, there is a great deal of heterogeneity of names at some geolocations, but many of the names are `similar' at precise addresses. There are also exact name matches `nearby,' highlighted in bold text.  Maps drawn using the tools of Ref. \cite{maptools}.}
\label{examples.fig}
\end{center}
\end{figure}

The fundamental difficulty that must be overcome in name disambiguation is the possibility of alternate or error-ridden spellings of names or addresses in the database.  Two examples of names requiring disambiguation are depicted in Fig. \ref{examples.fig}:  some of the names associated the National Institutes of Health (NIH) in Bethesda MD (assignee on $\sim 4,250$ patents), and some of the names associated with a prolific inventor, Rosa Maria Cuberes-Altisent, in the Barcelona area (inventing $\sim 80$ patents).  In both cases, the subsets of names shown in the white boxes of Fig. \ref{examples.fig} indicate the extreme heterogeneity in some inventor or assignee names, with at least 84 unique alternate spellings for the NIH and 24 unique alternate spellings for Rosa Maria Cuberes-Altisent.  Disambiguation of these names requires not only matching all of the possible variations in the spelling of the institution or individual, but also {\em{not}} matching other names that refer to different entities with similar names.  For example, the National Institute of Health of Japan has multiple patents under the name ``National Institute of Health,'' which we do not want to link to the NIH in Bethesda, and likewise an inventor with the name ``Jose Maria Duran-Altisent'' (inventing at least 4 patents in the Madrid area) should not be linked to Rosa Maria's identity despite the similarities in their names.  A wide range of methods of varying complexity have been generated to solve this disambiguation problem for authors of publications \cite{authority,authorCitation,authorCitation2,authorCollaboration,authorFeatures}, patent assignees \cite{assigneeDisambig1,assigneeDisambig2,hall2001nber}, and patent inventor names \cite{inventorDisambig,inventorDisambig1,disambigProblems} (with the disambiguation of Li et. al \cite{inventorDisambig} a recent and comprehensive result for the USPTO).  In the case of inventor disambiguation, these methods will generally compare pairs of names using the similarity of the text of the names as well as data regarding the assignees, patent citations, patent classes, and geographical information. 

The geographical information found in the USPTO typically suffers from low quality addresses, where less than 5\% of USPTO inventors complete the street field in their address on the patent (and city- or zipcode-level information is the highest resolution available).  At this level, geolocation can be used as one of many rough indicators of the similarity between two names when comparing them for disambiguation.  However, patents in the EPO or PCT databases are found to contain higher resolution addresses in a far greater fraction of cases (where a street number, street, city, state, and zip code are often all provided), which can provide much greater specificity when comparing inventors:  if two inventors have `similar' names {\em{and}} live at exactly the same address, it is far more likely they refer to the same person than if they had `similar' names and lived in the same general area.  The same state of affairs exists for assignee names, with the EPO and PCT addresses often having street-level information and the USPTO addresses tending to be of low quality.

\begin{figure}[htbp]
\begin{center}
\includegraphics[width=\textwidth]{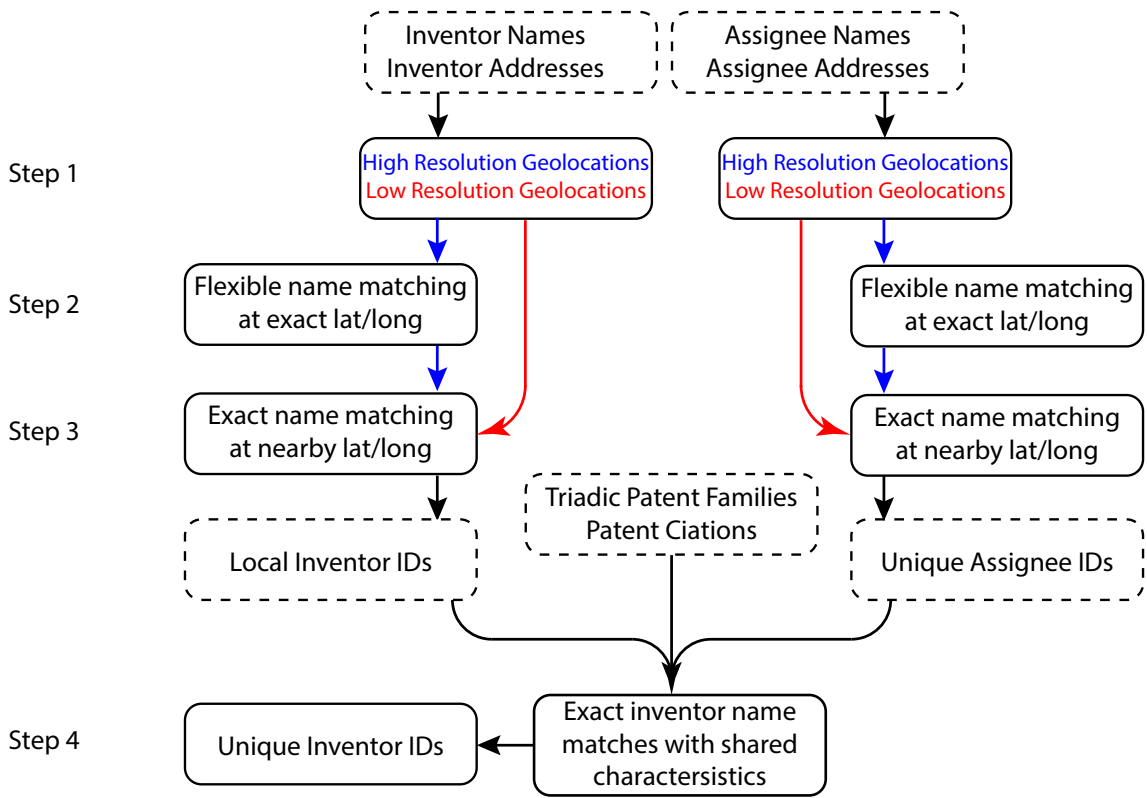}
\caption{Schematic diagram of the method.  Names and addresses are first geolocated (discussed in Sec. \ref{Geoloc.sec}), after which there is a search for `similar' names at exact high resolution lat/longs (the meaning of `similar' is discussed in Sec. \ref{assigneeExact.sec} for assignees and Sec. \ref{inventorExact.sec} for inventors).  Once similar names are clustered at each lat/long, `nearby' exact name matches are linked (described in Sec. \ref{nearby.sec}).  Exact name matches for inventors that are not `nearby' are checked for additional characteristics in common before a link is made (described in Sec. \ref{mobile.sec}).}
\label{schematic.fig}
\end{center}
\end{figure}

Our strategy for disambiguation will be to leverage these high precision addresses by flexibly matching names that are simultaneously found at exact, high-precision geolocations in any patent office, then to match `nearby' names that are exactly the same.  The general idea is sketched in Fig. \ref{examples.fig} for specific assignee and inventor names, and a schematic of the methodology is diagrammed in Fig. \ref{schematic.fig}.  A detailed summary of each step can be found in the appendix.  We first geolocate the assignees and inventor addresses for every patent in the three databases ($\sim 4$ million unique addresses) using Yahoo's YQL API \cite{yahoo,catini}, converting the text into likely latitude/longitude (lat/long) pairs and the quality of that geolocation (step 1 in Fig. \ref{schematic.fig}).  The quality returned by YQL generally indicates \cite{yahooquality} if the geolocation was resolved at the level of a point (e.g. ``55 Fruit Street, Boston, MA, USA''), line (e.g. ``Fruit Street, Boston, MA, USA''), zip code (e.g.  ``02114 MA USA''),  city (e.g. ``Boston, MA, USA''), state (e.g. ``MA, USA''), or country (e.g. ``USA'').  For every high-quality geolocation we attempt to flexibly match similar strings (step 2 in Fig. \ref{schematic.fig}), with specific examples in Fig. \ref{examples.fig}:  assignees with names involving `institute' or `health' at identical lat/longs are likely to be referring to the NIH, and inventors with names like `cuberes' and `rosa' at identical lat/longs are likely referring to Rosa Maria Cuberes-Altisent.  After this first round of disambiguation (which depends strongly on the existence of high-resolution geolocation information for the names) we search for exact name matches that were geolocated to a lat/long pair of {\em{any}} quality within 20km (step 3 in Fig. \ref{schematic.fig}).  In Fig. 1, we link the names at the indicated high-resolution geolocations due to the simultaneous name matchings of ``National Institutes of Health'' and ``Cuberes Altisent, Rosa''.   We also link occurrences of ``National Institutes of Health'' with the low-quality geolocations from addresses of ``Bethesda, MD, USA'' or ``Rockville, MD, USA'' that commonly occur in the USPTO, since those names are found within 20km of each other.   Having produced disambiguated assignee and inventor identifiers locally, we perform a final search for inventors that are mobile (where the same individual provides addresses more than 20km apart; step 4 in Fig. \ref{schematic.fig}).   Similarity between inventors of the same name beyond the 20km threshold is evaluated using additional indicators:  self-citations, shared co-assignees, shared co-inventors, and patent family overlap.  

Our approach produces a total of $\sim$800k unique assignee identifiers, $\sim $5.5M unique local inventor IDs, and links $\sim$800k mobile inventors worldwide into $\sim 360$k unique inventor IDs in the three patent offices, covering $\sim$9.1M patents.  Because our method is heavily dependent on high resolution geolocations of addresses, it is often important to distinguish between assignees or inventors that are linkable to high-quality addresses (e.g. the ``National Institutes of Health'' in ``Bethesda, MD, USA'' can be linked to the high-resolution geolocation in Fig. \ref{examples.fig}, despite it being a low resolution geolocation) and those that cannot (e.g. ``SONY CORP'' always provides the low-quality address ``Tokyo JP,'' and can never be linked to a high-resolution geolocation).  Names that {{cannot}} be linked to a high resolution address are incapable of being disambiguated for alternate spellings (since we only look for {\em{exact}} name matches when attempting to match names with low-resolution geolocations).  We find a total of $\sim$290k disambiguated assignees with at least one high-resolution address, $\sim$1.7M disambiguated inventors with at least one high-resolution address, and $\sim$275k mobile inventors merged into $\sim$ 190k high-resolution inventor IDs.  As shown in Table \ref{summaryTable.tab}, the coverage of high-resolution disambiguations is best in the EPO and worst in the USPTO, and the coverage of complete assignee disambiguation tends to be higher than that of inventor disambiguation.   These high resolution disambiguations provides complete coverage of every assignee and inventor on 3,583,475 patents worldwide (44\% in the EPO, 25\% in the PCT, and 31\% in the USPTO), and partial disambiguation (at least one assignee or inventor disambiguated with a high-resolution address) on $\sim$6.0M patents worldwide.  

%

\begin{table}[htdp]
\begin{center}
\begin{tabular}{|c|ccc|}
\hline
 & EPO & PCT & USPTO\\
 \hline
Patents with Assignees & 2.68 & 2.36 & 3.61\\
&&&\\
Patents with Assignees & 2.68 & 2.36 & 2.01\\
and addresses &&&\\
Patents with $\ge 1$ high-res&1.95 & 1.44 & 1.09\\
disambiguated ID &&&\\
Patents with all high-res& 1.91 & 1.39 & 1.07\\
disambiguated IDs & (71\%) & (52\%) & (30\%)\\
\hline
Patents with Inventors & 2.67 & 2.34 & 4.24\\
&&&\\
Patents with Inventors & 2.67 & 2.31 & 4.24\\
and addresses &&&\\
Patents with $\ge 1$ high-res& 1.99 & 1.44 & 1.98\\
disambiguated ID &&&\\
Patents with all high-res & 1.71 & 1.04 & 1.26\\
disambiguated IDs & (64\%) & (44\%) & (30\%)\\
\hline
\end{tabular}
\end{center}
\caption{Summary of the coverage of the disambiguation in the three patent offices (patent count in millions).  Listed are the number of patents having an assignee or inventor (note that these fields may be blank); the number of patents with any address information on the level of street, city, or zip code; the number of patents for which {\em{at least one}} of those entities is disambiguated with a link to a high resolution address; and the number of patents for which {\em{all}} of those names are disambiguated with a link to a high resolution address.  The percentages refer to the ratio of patents with a high-resolution disambiguation for every entity on the patent to the number of patents with at least one inventor or assignee name provided.}
\label{summaryTable.tab}
\end{table}%

%
%

\section{Benchmarks}

In order to benchmark the accuracy of our disambiguation, it is necessary to find a set of ground truth disambiguations of both patent assignees and patent inventors:  a manually curated subset of the patent data for which a correct disambiguation of the names has been performed.  Such a benchmark exists for $\sim 100$ USPTO inventors of $\sim 1300$ patents in the area of engineering and biochemistry that has been used as a golden standard in other inventor disambiguation methods \cite{inventorDisambig}.  In the case of assignees we are not aware of such a `gold standard' disambiguation, and we manually generated our own benchmark from a by-hand disambiguation of a small subset of patent assignees in the EPO and PCT data.  To create a benchmark of manageable size, we focused on assignees in the EPO and PCT in specific regions active in specific fields of research.  The OECD REGPAT database \cite{regpat} provides geolocation information on the level of NUTS3, and we generated a list of all assignees on patents assigned to names in Boston (having addresses in the NUTS3's US25017, US25025, or US25021) and Paris (having addresses in NUTS3's beginning with FR10).  These names include both regional assignees (those geolocated in the Boston or Paris areas) as well as external collaborators.  In order to reduce the number of names to disambiguate, we retained only assignees with at least one biopharma patent (defined as having an IPC classification which falls under the fields ``Pharmaceuticals'' or ``Biotechnology'' using the WIPO field-level aggregation of patent classes \cite{wipo}).  Manual matching of the names was often straightforward due to their clear similarity (e.g. ``ZENECA Pharma S.A.'' and ``ZENECA-PHARMA'' in the Paris area were given the same ID), but a web search was performed for somewhat similar names to see if there were any equivalent names in the data (e.g. ``Tufts Medical Center'' and ``New England Medical Center'' are synonymous in the Boston area, and were given the same ID).  This produced a final set of 1,295 disambiguated assignees (from 1,444 raw names) on 23,221 biopharma patents in the Boston area and 1,137 disambiguated assignees (from 1,311 raw names) on 18,645 biopharma patents in the Paris area.  

\begin{figure}[htbp]
\begin{center}
\includegraphics[width=.4\textwidth]{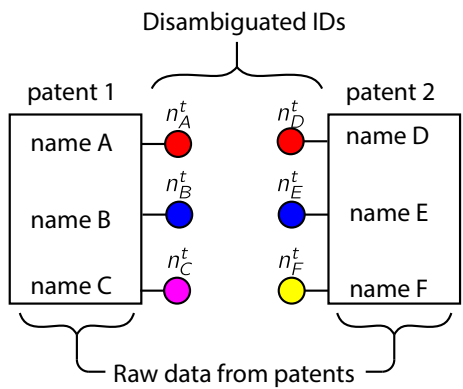}
\caption{Schematic of the pairwise measurements of precision and recall.  Each name is assigned a unique ID in the benchmark (indicated by the color of the circles) and a ID in the trial disambiguation (indicated by $\nv^t$).  Two true positives occur if $n_A^t=n_D^t$ and $n_B^t=n_E^t$.  Two false negatives would occur if $n_A^t\ne n_D^t$ and $n_B^t\ne n_E^t$.  A false positive occurs if $n_C^t=n_F^t$.}
\label{precisionRecall.fig}
\end{center}
\end{figure}

We will quantify the similarity between two disambiguations (clusterings of names into unique identifiers) a measure of precision and recall based on pairwise comparisons of the patents \cite{InfoRetrievalBook}.  For our manual benchmarks in Boston and Paris, where all assignee names on each patent have been disambiguated, it is straightforward to do a pairwise comparison of each patent.  The disambiguated name-to-ID's $\nv^b(p)=\{n_i^b(p)\}$ for the $i^{th}$ name on patent $p$ that can be compared to the disambiguated names in the trial disambiguation (with IDs $\nv^t(p)\{n_i^t(p)\}$ for the $i^{th}$ name).  For each pair of patents $p_1$ and $p_2$, we can determine the number of true positives, false positives, and false negatives between the trial and benchmark IDs by comparing the number of overlapping identifiers using the two disambiguations, as diagrammed in Fig. \ref{precisionRecall.fig}.  Defining $I_b(p_1,p_2)=|\{n_k^b(p_1)\cap n_k^b(p_2)|$ as the size of the intersection of the IDs in the benchmark and $I_t(p_1,p_2)=|\{n_k^t(p_1)\cap n_k^t(p_2)|$ the size of the intersection in the trial disambiguation, there are at most $TP(p_1,p_2)=\mbox{min}(I_b,I_t)$ IDs that agree in both partitions, $FN(p_1,p_2)=\mbox{max}(0,I_b-I_t)$ matches in the benchmark not seen in the trial, and $FP(p_1,p_2)=\mbox{max}(0,I_t-I_b)$ matches in the trial partition that don't match in the benchmark.  An estimate\footnote{Note that this method does not compare the positions of the name matches (only the number of matches), and thus neglects the possibility of a transposition of the IDs in the trial (e.g. if names A and E are incorrectly linked together and {\em{simultaneously}} names B and D are incorrectly linked together).  Due to the large number of IDs for both assignees and inventors, this type of error is expected to have a negligible effect.  This approximation can be relaxed, at an increased computational cost.} of the precision and recall for the trial is then
\begin{small}
\begin{eqnarray}
precision=\frac{\sum_{p_1\ne p_2} TP(p_1,p_2)}{\sum_{p_1\ne p_2} [TP(p_1,p_2)+FP(p_1,p_2)]}\qquad recall=\frac{\sum_{p_1\ne p_2} TP(p_1,p_2)}{\sum_{p_1\ne p_2} [TP(p_1,p_2)+FN(p_1,p_2)]}\label{precAndRecall}
\end{eqnarray}
\end{small}
Using the definitions in Eq. \ref{precAndRecall}, if all matches found in the trial are also found in the benchmark, the trial disambiguation will have high precision; and if all matches in the benchmark are found in the trial, the trial disambiguation will have high recall.

\begin{table}[htdp]
\begin{center}
\begin{tabular}{|c|cc|cc|}
\hline
trial & precision & precision & recall & recall\\
& (Boston) & (Paris) & (Boston) & (Paris)\\
\hline
No Disambig. & 1.000 & 0.999 & {\bf{0.734}} & {\bf{0.519}}\\
\hline
OECD HAN & 0.991 & 0.966 & {\bf{0.837}} & {\bf{0.674}}\\
\hline
Our method & 0.993 & 0.996 & {\bf{0.985}} & {\bf{0.972}}\\
\hline
\end{tabular}
\end{center}
\caption{Comparison of three assignee disambiguations to the benchmarks in the Boston and Paris Regions:  a manual disambiguation of all assignees on EPO or PCT patents with at least one assignee in the Boston or Paris areas.  All methods produce high precision, because similar names in the same metropolitan area are likely to refer to the same institution.  The HAN harmonization provides a modest improvement over the recall of the raw names in both Boston and Paris, and our algorithm shows a more significant increase in recall than the HAN identifiers.}
\label{assigneeBenchmark.tab}
\end{table}%

We compare our algorithm's disambiguation of assignee names with the manual benchmark in the Boston and Paris area in Table \ref{assigneeBenchmark.tab}, alongside undisambiguated names (after conversion to lowercase and dropping all punctuation) and the OECD HAN \cite{han} name harmonization (which accounts for synonyms such as ``Co'' and ``Company'' in assignee names).  We are not aware of any other freely available disambiguation of the OECD assignee names that we can use as an additional comparison.  In Table \ref{assigneeBenchmark.tab}, we see all trial disambiguations have high precision because the benchmark is biased towards institutions are in the same region, where similar names for assignees are almost certain to refer to the same institution.  For example, if all patents worldwide were included in the benchmark, a potential error could occur because both the Massachusetts General Hospital and the Rochester General Hospital (in Henrietta, New York) are sometimes referred to as ``The General Hospital Corp,'' and the identical names would reduce the precisions for both the undisambiguated and HAN Harmonized trials in Table \ref{assigneeBenchmark.tab}.  
However, because our benchmark specifically excludes any assignees that do not collaborate with a Boston- or Paris-area assignee, we greatly reduce the possibility of name conflicts (since distinct institutions in the same region are likely to distinguish themselves by differing names.), which leads to a high precision using each method.   The recall of the various trials shows a greater variation in both Boston and Paris, with the undisambiguated names performing worst, the HAN harmonization a marginal improvement in precision over the raw names, and our method showing a significant improvement over both.  The lower overall recall in the Paris area is due to many alternate spellings from the presence or absence of accents on words, a complication not typically present in the Boston area.  Note that the names in the Paris area involve terms in French (e.g. ``societ\'e aononyme'' is a commonly occurring substring of assignee names), and the high precision of our disambiguation in the Paris area suggests that the algorithm is robust to variations in language.

\begin{table}[htdp]
\begin{center}
\begin{tabular}{|c|c|c|}
\hline
trial & recall & recall\\
& (all patents in & (patents restricted to\\
&  benchmark) &  high-res inventors)\\
\hline
No Disambig. & 0.743 & 0.799\\
Our IDs & 0.779 & 0.860\\
Li IDs & 0.887 & 0.885\\
\hline
\end{tabular}
\end{center}
\caption{Recall for the three different disambiguation approaches on the inventor benchmark \cite{inventorDisambig}.  In the first column, the disambiguated IDs are restricted to only inventors with the same first and last names as those occurring in the benchmark, and only patents in the benchmark are included.  Precision for all measures is identically 1, expected since patents outside of the benchmark are excluded.  In the second column, we exclude patents with at least one inventor who had no link to a high-resolution address using our algorithm, and thus focuses on patents were we expect our algorithm to perform well.}
\label{inventorBenchRecall.tab}
\end{table}%

Assignees generally have much more complex names and the many alternate spellings are a primary source of difficulty, while inventor names generally have a lower propensity for alternate spellings but have the potential for mobility.  Due to the significant differences between the disambiguation of assignees and inventors, it is thus important to evaluate the accuracy of our inventor disambiguation as well.  To evaluate the accuracy of our inventor disambiguation, we compare it to to a benchmark \cite{inventorDisambig} based on the examination of inventor CVs and in-person interviews.  Since not every inventor's CV was included in the benchmark, the collaborators of inventors in the benchmark are generally not disambiguated.  We therefore drop inventor IDs generated using our algorithm that did not have an identical inventor name as appearing in the benchmark when evaluating the similarity through precision or recall.  These inexact name matches are assumed\footnote{In a small fraction of the cases, differences between the raw data and the patent data prevented an exact linking of the disambiguated data (e.g. the inventor names ``Tsu Jae King Liu'' vs ``Tsu Jae King'' in the data and benchmark, respectively, on patent US7807523).  These patents are not included in the recall statistics.} to be collaborators of the inventors in the benchmark rather than the benchmarked inventors themselves, and inclusion of these missing inventors could cause excessive false negatives or spurious false positives during the pairwise comparison.  The recall for this comparison is shown in the second column of Table \ref{inventorBenchRecall.tab} using the undisambiguated names (again after removing punctuation and conversion to lowercase), our algorithm, and the USPTO inventor disambiguation of Li et al \cite{inventorDisambig}.  We see that, unlike the strong results in Table \ref{assigneeBenchmark.tab}, our algorithm has only a very modest improvement in recall over the undisambiguated names.  The tiny improvement in recall in the first column of Table\ref{inventorBenchRecall.tab} is due primarily to the poor quality of the inventor addresses in the USPTO:  many inventors provide only general geolocation information, and our algorithm (depending strongly on precise addresses to correct for variations in names) will not be able to perform any meaningful disambiguation at all on these inventors.  If we restrict the comparison {\em{only}} to patents for which the inventors are matched to at least one high-resolution geolocation\footnote{For each patent, we keep patents for which the inventor could be associated with at least one high-resolution address {\em{in the entire dataset}} after disambiguation, not just on the patent under consideration.  Patents for which the inventors provided a low-quality address on {\em{every}} patent in the database are excluded}, the recall increases significantly (as shown in the third column of Table. \ref{inventorBenchRecall.tab}).  Our algorithm is comparable to the disambiguation of Li et al \cite{inventorDisambig} if we restrict only to patents for which we the method can successfully use the geolocation to correct for noise in the names using this benchmark.  In general, we see that the recall of all of the  methods do not differ too much from one another on this benchmark, indicating the intrinsic difficulty of accurate inventor disambiguation (particularly for small samples such as this).

Splitting and lumping \cite{inventorDisambig} are two alternate statistics that have been computed when benchmarking inventors, which respectively estimate the number of patents or IDs that are missing in the trial disambiguation in comparison to the benchmark or the number of patents or IDs that are added in the trial disambiguation compared to the benchmark.  To compute the split and lumped patents for a specific inventor $i$ in the benchmark, we match him or her to the disambiguated ID in the trial with the greatest number of patents in common with inventor $i$, denoted $m_i$ (in the case of a tie, we then choose the one with the fewest patents {\em{not}} in common).  Defining $\Pv_i^b$ the set of patents invented by $i$ in the benchmark and $\Pv_{m_i}^t$ the set of patents invented by the ID $m_i$ in the trial disambiguation, the number of split patents is $s_i=|\Pv_i^b\backslash \Pv_{m_i}^t|$ (the patents found in the benchmark missing in the trial) and number of lumps is $l_i=|\Pv_{m_i}^t\backslash \Pv_i^t|$ (the patents found in the trial absent in the benchmark).  The splitting and lumping statistics are defined as\footnote{Note that this definition may vary slightly from others, as it is possible to focus on the number of inventors grouped together or separately rather than the number of patents grouped together or separately.}
\begin{eqnarray}
splitting=\frac{\sum_i s_i}{\sum_i |\Pv_i^b|}\qquad lumping=\frac{\sum_i l_i}{\sum_i |\Pv_i^b|}.\label{splitlump}
\end{eqnarray}  
In table \ref{splittingLumping.tab}, we apply this comparison to the verified inventor benchmark (those for which the authors of \cite{inventorDisambig} were able to confirm their inventorship in person) restricted to patents where all inventors are linked to a high-resolution geolocations\footnote{Our IDs perform relatively poorly if we do not excluded low-quality disambiguations.  The results of Li performs somewhat worse if we include all patents in the benchmark, rather than restricting to the verified names.}.  If all patents are considered or algorithm is only a modest improvement over the use of the undisambiguated names (data not shown), but if we restrict our focus on patents with inventors having high-resolution geolocations our algorithm is comparable to the statistics for the disambiguation of Li.

\begin{table}[htdp]
\begin{center}
\begin{tabular}{|c|c|c|}
\hline
data & splitting & lumping\\
& (high-res and & (high-res and\\
& verified patents) & verified patents)\\
\hline
No Disambig. & 0.384 & 0.012 \\
Our IDs & 0.050 & 0.038\\
Li IDs & 0.067 & 0.047\\
\hline
\end{tabular}
\end{center}
\caption{Splitting and lumping statistics for the various inventor disambiguations, restricted to the {\em{verified}} inventor benchmark \cite{inventorDisambig} (inventors with CVs that also responded to the questionnaire confirming they were involved in the listed patents) that could also be linked to high-resolution geolocations, and includes 659 patents.  Smaller numbers are better.  Our algorithm and the disambiguation of Li et al \cite{inventorDisambig} produce comparable results, with a drastic decrease in the splitting in comparison to the undisambiguated names, with a moderate increase in the lumping.}
\label{splittingLumping.tab}
\end{table}%

\section{Conclusions}

In this paper, we have described a new algorithm for disambiguating the assignees and inventor names for a significant fraction of the patent data from the EPO, PCT, and USPTO simultaneously.  Our approach focuses heavily on high-resolution geolocation of assignee and inventor addresses, determined by uploading all of the address information provided in the databases to the Yahoo Query Language \cite{yahoo}.  This conversion from text to latitude/longitude pairs not only provides the possibility of precisely locating inventors and assignees on a map, but also acts as a disambiguation of the (often error prone) address fields.  We use the high-resolution addresses to search for similar names at the same address using a very flexible string matching algorithm that are geolocated to {\em{identical}} latitude/longitude pairs, then search for {\em{identical}} names within a radius of 20km.  The patents linked to assignees and inventors that can be matched to at least one high-resolution geolocation cover $\sim $40\% of the patents worldwide (with the best coverage in the EPO and the worst in the USPTO). The variability in coverage is due to the fact that some patent offices (the USPTO in particular) do not always include high quality addresses, and the approach we implement here would be impossible without treating all three patent offices simultaneously to capture as many high resolution geolocations as possible.  We show that our method is able to accurately disambiguate both assignees and inventors on two benchmarks using a manual disambiguation of assignee names in the Boston and Paris areas for patents in the EPO and PCT, and a widely used benchmark of $\sim100$ inventors \cite{inventorDisambig} from the USPTO.  Our method had high precision and recall for the assignees in both regions considered, and that our disambiguation of the inventors was comparable to the disambiguation of Li et al\cite{inventorDisambig} when restricted to fully-disambiguated patents.  

The disambiguation resulting from our method is immediately useful in a variety of contexts, but we intend to explore further improvements to expand the accuracy and (more importantly) the coverage of the disambiguation of both assignees and inventors.  There are almost certainly minor improvements that can be made in the matching of `similar' names at a specified lat/long (where a machine learning techniques for assigning similarity may be more robust to regional variations in names than our hard-threshold approach), but a more significant improvement that could be made is the linking of names with low-quality addresses with high-quality disambiguated names.  Our current method depends on an {\em{exact}} matching of assignee or inventor names, and the inclusion of names without an associated high-resolution address link tends to significantly lower the power of our method (as was shown in Table \ref{inventorBenchRecall.tab}).  Some specific examples of improvements we are currently exploring are
\begin{enumerate}
\item{Better handling of corporate synonyms as is done in the HAN \cite{han} harmonization (where e.g. ``AG'' and ``Aktiengesellschaft'' are treated as identical when string matching). This would be an update to Step 2 in Fig. \ref{schematic.fig}.}
\item{Link names across patent families in a manner similar to that of high-resolution addresses.  For example, if ``Sony Corporation'' is found on one patent and ``Sony Electronics'' is found on another patent {\em{in the same patent family}} with a nearby address, link those two names.  This would be an update to step 3 of Fig. \ref{schematic.fig}.  Expansion of this idea to other similarity indicators (shared inventors, for example) will also be explored.}
\item{Rather than requiring inventors or assignees to have an {\em{exact}} name match nearby, searching for names that have a small (but nonzero) Levenshtein distance (in the entire string, not in individual words) with other names nearby.  This would be an update to step 3 of Fig. \ref{schematic.fig}.}
\item{Rather than requiring mobile inventors to have exact name matches in multiple locations, search for names that have a small (but nonzero) Levenshtein distance that share assignees, coinventorships, citations, or patent families.  This would be an update to step 4 of Fig. \ref{schematic.fig}}
\end{enumerate}
Item 2 in this list could also be used to link assignees that provided no address information whatsoever (occurring on $\sim 1$M USPTO patents) to disambiguated IDs.  Another major source of improvement would be to reduce the number of low resolution geolocations, by seeking out alternate geolocation techniques or APIs.  Many addresses have more than street level information but are geolocated at a low resolution due to YQL's inability to match that address to a specific point.  This is a significant problem outside of the US and western Europe, with Japan particularly affected:  of the $\sim100k$ address containing the word `Tokyo,' 0.6\% produced a high-resolution geolocation but 82\% are found on the neighborhood level or better (meaning they contained more information than simply a city name).   In addition to these improvements, more refined machine learning techniques \cite{authority,inventorDisambig} that avoid some of the hard thresholds we have introduced in the algorithm would likely increase its robustness in many places.  The implementation of these modifications and their effect on the accuracy and coverage of the disambiguation in all three patent offices will be discussed in future publications.

\section{Data Release}

Our data is being made freely available for noncommerical use, and will be released after a successful peer review evaluation.  The data will be located at\\
https://github.com/gmorriso/PatentDisambigData \\
when available.  The data that will be provided includes
\begin{itemize}
\item{A patent-to-entity list of all 3.6 million patents to the unique assignee and inventor IDs, for which all entities can be linked to a high-resolution address.}
\item{A correspondence between each assignee and inventor ID and the various names that were disambiguated using that ID.}
\item{The manual institution disambiguation in the Boston and Paris areas.}
\end{itemize}
For questions prior to release of the data, please contact greg.morrison$@$imtlucca.it.

\noindent \textbf{Acknowledgements:} 
We acknowledge support from the Crisis Lab Project (MIUR, Italy) and from the CERM Foundation (Switch Project).  We are also grateful for the assistance of Anna Horodok on some of the manual disambiguation and for many useful conversations with Orion Penner. 


\begin{thebibliography}{10}

\bibitem{spillovers}
J~Owen-Smith and WW~Powell.
\newblock Knowledge networks as channels and conduits: The effects of
  spillovers in the boston biotechnology community.
\newblock {\em Organization Science}, 15:5, 2004.

\bibitem{miguelez}
E~Migu{\'e}lez and and R~Moreno
\newblock Knowledge flows and the absorptive capacity of regions.
\newblock {\em Research Policy}, 44:4, 2015.

\bibitem{marx2007noncompetes}
M~Marx, D~Strumsky, and L~Fleming.
\newblock {\em Noncompetes and inventor mobility: Specialists, stars, and the
  Michigan experiment}.
\newblock Division of Research, Harvard Business School, 2007.

\bibitem{mobility}
C~Franzoni, G~Scellato, and P~Stephan.
\newblock Foreign born scientists: Mobility patterns for sixteen countries.
\newblock {\em Nature Biotech}, 30:1250, 2012.

\bibitem{bordercollab}
A~Chessa, A~Morescalchi, F~Pammolli, O~Penner, AM~Petersen, and M~Riccaboni.
\newblock Is europe evolving toward an integrated research area?
\newblock {\em Science}, 339:650, 2013.

\bibitem{mansci}
J~Owen-Smith, M~Riccaboni, F~Pammolli, and WW~Powell. 
\newblock A comparison of US and European university-industry relations in the life sciences.
\newblock {\em Management science}, 48:1, 2002.

\bibitem{riccaboni}
M~Riccaboni, WW~Powell, F~Pammolli, and J~Owen-Smith.
\newblock Public research and industrial innovation: a comparison of US and European innovation systems in the life sciences.
\newblock {\em Science and Innovation. Rethinking the Rationales for Funding and Governance}, Cheltenham: Edward Elgar, 2003.

\bibitem{morescalchi}
A~Morescalchi, F~Pammolli, O~Penner, A~Petersen, and M~Riccaboni.
\newblock The evolution of networks of innovators within and across borders: Evidence from patent data.
\newblock {\em Research Policy}, 44:3, 2015.

\bibitem{morrisonjcn}
G~Morrison, G~Eleftherios, F~Pammolli, and M~Riccaboni.
\newblock Border sensitive centrality in global patent citation networks.
\newblock {\em Journal of Complex Networks}, 2:4, 2014.

\bibitem{youn2015invention}
H~Youn, D~Strumsky, LMA Bettencourt, and J~Lobo.
\newblock Invention as a combinatorial process: evidence from us patents.
\newblock {\em J of The Royal Soc Interface}, 12:20150272, 2015.

\bibitem{authority}
VI~Torvik and NR~Smalheiser.
\newblock Author name disambiguation in medline.
\newblock {\em ACM Trans Knowl Discov Data}, 3:11, 2009.

\bibitem{inventorDisambig}
GC~Li et~al.
\newblock Disambiguation and co-authorship networks of the us patent inventor
  database (1975Ð2010).
\newblock {\em Research Policy}, 43:941, 2014.

\bibitem{assigneeDisambig2}
P~Cuxac, JC~Lamirel, and V~Bonvallot.
\newblock Efficient supervised and semi-supervised approaches for affiliations
  disambiguation.
\newblock {\em Scientometrics}, 97:47, 2013.

\bibitem{inventorDisambig1}
M~Pezzoni, F~Lissoni and G~Tarasconi.
\newblock How to kill inventors: testing the Massacrator{\copyright} algorithm for inventor disambiguation.
\newblock {\em Scientometrics}, 101:1, 2014.

\bibitem{maptools}
Free~Map Tools.
\newblock {\em http://www.freemaptools.com/radius-around-point.htm}, 2015.

\bibitem{authorCitation}
H~Han, L~Giles, H~Zha, C~Li, and K~Tsioutsiouliklis.
\newblock Two supervised learning approaches for name disambiguation in author
  citations.
\newblock In {\em Proceedings of the 2004 Joint ACM/IEEE Conference on Digital
  Libraries, 2004}, page 296, 2004.

\bibitem{authorCitation2}
L~Tang and J~Walsh.
\newblock Bibliometric fingerprints: name disambiguation based on approximate
  structure equivalence of cognitive maps.
\newblock {\em Scientometrics}, 84:763, 2010.

\bibitem{authorCollaboration}
IS~Kang et~al.
\newblock On co-authorship for author disambiguation.
\newblock {\em Inf. Process. and Manag.}, 45:84, 2009.

\bibitem{authorFeatures}
M~Levin, S~Krawczyk, S~Bethard, and D~Jurafsky.
\newblock Citation-based bootstrapping for large-scale author disambiguation.
\newblock {\em J Am Soc Info Sci and Tech}, 63:1030, 2012.

\bibitem{assigneeDisambig1}
B~Balsmeier et~al.
\newblock Automated disambiguation of us patent grants and applications.
\newblock {\em {http://tinyurl.com/BBalsmeier}}, 2015.

\bibitem{hall2001nber}
BH~Hall, AB~Jaffe, and M~Trajtenberg.
\newblock The nber patent citation data file: Lessons, insights and
  methodological tools.
\newblock Technical report, National Bureau of Economic Research, 2001.

\bibitem{disambigProblems}
SL~Ventura, R~Nugent, and ERH Fuchs.
\newblock Methods matter: Revamping inventor disambiguation algorithms with
  classification models and labeled inventor records.
\newblock {\em SSRN eLibrary}, 2012.

\bibitem{yahoo}
Yahoo Inc.
\newblock {YQL} web service {URL}s.
\newblock {\em https://developer.yahoo.com/yql/guide/yql\_url.html}, 2015.

\bibitem{catini}
R~Catini, D~Karamshuk, O~Penner, M~Riccaboni
\newblock Identifying geographic clusters: A network analytic approach.
\newblock {\em Research Policy}, 44:9, 2015.

\bibitem{yahooquality}
Yahoo Inc.
\newblock Supported response formats.
\newblock {\em
  https://developer.yahoo.com/boss/geo/docs/\\supported\_responses.html\#address-quality},
  2015.

\bibitem{regpat}
{OECD} {REGPAT} database.
\newblock July 2014.

\bibitem{wipo}
U~Schmoch.
\newblock Concept of a technology classification for country comparisons.
\newblock {\em Final Report to the World Intellectial Property Office (WIPO)},
  2008.

\bibitem{InfoRetrievalBook}
CD~Manning, P~Raghavan, and H~Sch{\"u}tze.
\newblock {\em Introduction to information retrieval}.
\newblock Cambridge university press Cambridge, 2008.

\bibitem{han}
{OECD} {HAN} database.
\newblock July 2014.

\bibitem{regpatold}
{OECD} {REGPAT} database.
\newblock January 2014.

\bibitem{openstreetmap}
B~Quinion.
\newblock Open source search based on {O}pen{S}treet{M}ap data.
\newblock {\em https://github.com/twain47/Nominatim}, 2015.

\bibitem{googleplaces}
Google Inc.
\newblock The google maps geocoding {API}.
\newblock {\em
  https://developers.google.com/maps/documentation/geocoding/intro}, 2015.

\bibitem{geonames}
Geo{N}ames.
\newblock {\em http://www.geonames.org}, 2013.

\end{thebibliography}


\appendix

\section{Methodology}

\subsection{Geolocation}
\label{Geoloc.sec}

Our disambiguation relies heavily on the geolocation of inventor and assignees using the addresses provided in the patent data.  The names and addresses for each patent were extracted from the OECD Regpat January 2014 database \cite{regpatold} for the EPO and PCT patents, and from the patent database provided by Li et al \cite{inventorDisambig} for the USPTO patents.  The addresses are uploaded to yahoo's YQL \cite{yahoo} API (with a UTF-8 encoding of the address; no additional cleaning performed), with a JSON response returned by the server containing a great deal of information about the geolocation(s) of that address.  If the address was successfully located by YQL, we extracted all latitude and longitude data along with the quality of that geolocation \cite{yahooquality}, where the quality is an assessment of how precise the lat/long is (e.g. street level vs. city level).  Note that in principle other geolocation APIs could be used (e.g. OpenStreetMap \cite{openstreetmap} or Google places \cite{googleplaces}), but YQL was chosen due to its familiarity and ease of use.  After the geolocation was complete on the January data, we acquired the July 2014 OECD Regpat database \cite{regpat}.  The geolcations extracted from the January database addresses were applied to the July database without modification, so some addresses from the newer data release may be missing.  

The number of patents with geolocations are listed in Table \ref{geocount.tab}, with high coverage of the geolocation for all inventors and all assignees {\em{except}} for the USPTO.  1,847,909 USPTO patents have absolutely no assignee address information (no assignee provided, or no information about that assignee listed).  Only 2,168,220 USPTO patents have assignee address information of any kind in the database (and $\sim$2M have address at the resolution of city or better).  As our algorithm depends strongly on geolocation, only these $\sim$2M USPTO patents will have even a chance of acquiring a disambiguated assignee.  Over 95\% of the patents filed in each office have at least one inventor geolocation, and coverage is good in the EPO and PCT for assignees as well.  Only $\sim$90\% of the assignee addresses for the USPTO result in a geolocation, due to the relatively low quality of assignee addresses found in the data.

\begin{table}[htdp]
\begin{center}
\begin{tabular}{|c|ccc|}
\hline
& EPO & PCT & USPTO\\
\hline
Patents in the database & 2.67 & 2.37 & 4.14\\
Patents with any geolocation(s) & 2.64 & 2.30 & 4.14\\
Patents with inventor geolocation(s) & 2.54 & 2.18 & 4.06\\
Patents with assignee geolocation(s) & 2.58 & 2.24 & 1.94\\
\hline
\end{tabular}
\end{center}
\label{geocount.tab}
\caption{Patents in the database (numbers in millions of patents), and the address information included in them.  Here, `geolocations' refer to address information uploaded to YQL that returned at least one non-empty response.  Empty responses are likely due to missing addresses (particularly due to the USPTO, where many assignees addresses are not provided).  }
\end{table}%

%
%
%
%
%
%
%
%

\subsection{Assignee name matching}
\label{assigneeExact.sec}

Assignee disambiguation can be extremely difficult due to the large number of alternate spellings of assignee names.  An example of a difficult disambiguation is shown in Table \ref{assigneeExample.tab} for alternate naming for the National Institutes of Health (NIH) in the Rockville / Bethesda areas of Maryland, USA, which is an agency of the Dept. of Health and Human services.  Patents produced by the NIH may have assignee names that solely include references to the Department to which it reports, or completely exclude the Department, or mention the Department in conjunction with the Institutes.  While this extreme variability makes disambiguation difficult, we note that each of the addresses in this table are geolocated to the same lat/long with high quality using YQL.  The geolocation is thus providing two services in this respect:  in addition to locating the specific place where the institution is located, it is providing a robust disambiguation of the addresses.  The similarity between the names in Table \ref{assigneeExample.tab} in conjunction with their precise geolocation to {\em{identical}} points certainly suggests that these names likely refer to the same entity, and in general one has good reason to believe that ``similar names'' found at identical high precision street address are likely to refer to the same institution.   Our algorithm for the first step of disambiguation is thus to block all names according to their high-resolution geolocations and search for ``similar names.''  Names that are ``similar'' are linked to one another as referring to the same disambiguated institutions.

\begin{table}[htdp]
\begin{center}
\begin{small}
\begin{tabular}{|c|c|c|}
\hline
patent & assignee name & assignee address\\
\hline
EP1807440 & Department of Health and Human Services & 6011 {\bf{Excecutive}} Boulevard,\\
& & Rockville MD 20852\\
\hline
EP2019710 & National {\bf{Institute}} of Health & Office of Technology Transfer\\
& & 6011 Executive Boulevard, Suite 325,\\
& & Rockville MD 20852-3804\\
\hline
EP1361886 & The Gov. of USA, as represented by the  & Office of Technology Transfer,\\
& Secretary, {\bf{Dept.}} of Health and Human  &6011 Executive Boulevard, Suite 325,\\
& services, National Institutes of Health & Rockville, MD 20852\\
\hline
\end{tabular}
\end{small}
\end{center}
\label{assigneeExample.tab}
\caption{Examples of three patent assignees that are ``similar'' but difficult to disambiguate.  A few misspellings or abbreviations are bold-faced.  Each of them are geolocated to at least one identical high-resolution latitude/longitudes ($39.048843,-77.120419$ with a quality rating of 87, specifically).  There is significant overlap between the words found in the assignee name in EP1361886 and the words in the assignees of either EP2019710 and EP1807440, suggesting these names should be matched. }
\end{table}%

In order to determine if two names are similar, it is useful to build a list of `common' and `rare' words (an algorithmically-defined distinction between common and rare has been useful in a variety of methods \cite{inventorDisambig}).  A dictionary of `common' words (e.g. `hospital' and `institute') that was generated by hand  (and passed through google translate in a few languages) is read into memory, as well as a dictionary of location names (e.g. `Boston') provided by GeoNames \cite{geonames} and all first and last names occurring in the inventors found in our databases.  All words on any of these lists are treated as `common' in the disambiguation algorithm.   A few words in the common list (e.g. ``Company'') were manually selected as completely uninformative, where the inclusion of these words caused many spurious incorrect matchings, and are removed from the names before matching.  For the manually curated common words, a misspelling dictionary is also constructed by (a) producing all deletion errors possible by deleting each character in turn, and (b) by producing every permutation error possible by swapping the order of every character in the name.  These permutations and deletions are {\em{not}} applied to the full set of inventor names or geographic names.  In order to check to see if a name is `common' within one error, we check to see if it is found in the common word list, or if it {\em{or any single deletion substring}} are found in the misspelling dictionary.  During assignee disambiguation, words that are on the common list or within one deletion from the misspelling dictionary are treated as `common,' otherwise they are treated as `rare.'  All of the assignee names are processed as described in Sec. \ref{string.sec}.  

To perform the first step of disambiguation, we search for similar names at each high-resolution geolocation.  In this draft, ``high-resolution'' refers to any geolocation with a YQL quality of at least 70, generally corresponding to a geolocation with street-line level accuracy or above.   Geolocations with an accuracy on the level of zip-codes, cities, or higher are ignored in this first step of the disambiguation process.  
The algorithm then iterates over each high-resolution lat/long, and all pairs of names at that location are compared to see whether they are similar:
\begin{enumerate}
\item{If any two `rare' words are within one edit distance of one another in both names, the names are a mach.  If all rare words in the first name are more than one edit distance from all words in the second name, proceed to step 2.  For example, ``Harvard University'' and ``Harvard College'' are matched in this step because `harvard' is a rare word.}
\item{If the list of `common' words in both names have at least two elements in common, the names are a match (if one of the names consists of exactly one word [e.g. ``Apple''] and that word is `common', only one shared `common' word is required).  If there is not a sufficient overlap between `common' words found in the names, proceed to step 3.  For example, ``The General Hospital Corporation'' and ``Massachusetts General Hospital'' are matched in this step, because `general' and `hospital' are common words.}
\item{Check to see if the first name contains an acronym found in the second name.  For each word in the first name, break it into individual letters and see if there is a subset of words in the second name (preserving the ordering) that all {begin} with those letters.  If no match is found, check for acronyms in the second name.  If no acronym is found at this step, we assume the names are distinct.  For example, ``The Massachusetts Institute of Technology'' and ``MIT'' are matched in this step, since `m', `i', and `t' are found sequentially in the first letters of words in the first name. }
\end{enumerate}

Having linked all of the names at each lat/long, every name/geolocation pair is assigned a unique identifier, with any names matched in the algorithm assigned the same identifier.  Beginning with the 352,393 name/high-resolution geolocation pairs found in the raw data after geolocation and name cleaning, there are 329,079 unique identifiers produced using this algorithm.  We note that this is a rather modest reduction in the number of name/geolocation pairs, and may appear to have done very little.  However, this step performs the essential service of cleaning very noisy names, of particular importance for large institutions.  These are the clustered names that are passed to the neighborhood search described in Sec. \ref{nearby.sec}.

\subsection{Inventor name matching}
\label{inventorExact.sec}

Inventor names generally have less variability in their structure than assignee names, where usually there is a `last name' (typically the first word in the name), a `first name' (typically the second word in the name) and finally various `middle names.'  In reality, the first and middle names are sometimes interchangeable, a two-word last name may be separated, and additional titles or company names may be added to the name (see Table \ref{inventorExample.tab} for two examples).   In order to overcome these types of errors, we adapt our word-based matching of names at the same high-resolution geolocation that was used to disambiguate assignees.  A manual inspection shows that these errors tend to be far more common in the EPO and PCT than in the USPTO (although it is difficult to quantify the rate of any particular error type without an accurate disambiguation in hand), but the USPTO tends to have far lower accuracy in the geolocations as well:  there are $\sim 4.5$M name/high-res geolocations in the combined EPO and PCT databases, while there are only $\sim$0.2M in the USPTO.  

\begin{table}[htdp]
\begin{center}
\begin{small}
\begin{tabular}{|c|c|c|}
\hline
patent & inventor name & inventor address\\
\hline
EP2340782 & GOMES DA CUNHA PONCIANO, & Av. Ipiranga 55 Centro,CEP: \\
&  Jos\'e Ant\^onio & 25685-250 Petr\'opolis, RJ\\
\hline
EP2386338 & GOMES, Jos\'e Antonio, da& Av. Ipiranga 55, Centro,\\
&  Cunha, Ponciano & Petr\'opolis - RJ, Cep: 25685-250\\
\hline
\hline
EP1247533 & Howard, Jr, Harry Ralph & Pfizer Golbal Res. and Dev., \\
& & Eastern Point Road,\\
& & Groton, Connecticut 06340\\
\hline
EP1220831 & HOWARD, Harry Ralph, & and Development Eastern Point Road\\
& Jr. Pfizer Global Reasearch & Groton, CT 06340\\
\hline
\end{tabular}
\end{small}
\end{center}
\label{inventorExample.tab}
\caption{Example of inventor names with a variety of errors.  The first two names refer to the same person, but the person's last name is split in the second occurrence (note also the missing accent in Antonio).  In the second example, `Jr' is put in the position of the first name in one instance and a portion of the address field is added to the inventor's name in another instance.  Both of these are geolocated to the same position, and a flexible matching can be performed.  }
\end{table}%

In order to disambiguate the inventor names, each name at a high-res geolocation is processed as described in Sec. \ref{string.sec}.  A search is performed for the strings ``c/o'' and ``c/-''; any words following this substring are forbidden from being matched at that geolocation under the assumption it refers to the assignee (matching due to that string is not forbidden at other geolocations).  In each name, it is assumed that the first two words in the name correspond to the last and first names, respectively, and we check to see if these assumed first and last names are found {\em{anywhere}} in the name we are comparing.  
\begin{enumerate}
\item{Check to see if the `last name' is found in the name we are comparing to.  If the word has more than three characters, allow for a difference of one edit.  We keep track of how often the `last name' was a match to the first word in the compared name.}
\item{If the `last name' was found in step 1, check to see if the `first name' is found in the compared name.  If there was {\em{perfect}} agreement in the last name, permit the `first name' match to differ by one edit distance.  If there was {\em{not perfect}} agreement in the `last name,' require perfect agreement with the `first name' (max. edit distance of 0).}
\end{enumerate}
For each pair of names, we perform this check using the first two words of each in our search.  If we find a `last name' $+$ `first name' match in between either of the names being compared at the same high-res geolocation, we link the names as referring to the same individual.  

This algorithm is fairly robust and able to disambiguate the names of a majority of high res geolocations without difficulty.  However, significant errors can occur due to the confluence of two events in the data:  (1) a large number of inventors using the same address and (2) culturally common `middle names.'  For example, the address ``Prof. Holstlaan 6,NL-5656 AA Eindhoven'' is used as an inventor address for over 34k EPO patents, and at that precise address there are 743 unique inventor names that have a middle name ``Maria.''  In such cases, last names that are one edit distance away from common last names will cause an overwhelming number of incorrect links between names (e.g. there is a person named ``Marra, Johannes'' at the same address, with a last name one edit distance from ``Maria'').  To prevent these huge errors, we perform a pruning step, and unlink pairs of names if the `last name' that caused the link was matched to a non-`last name' more than twice as often as it was matched to a `last name.'  This final post-processing step removes linked names when the `last name' is overwhelmingly matched to a `first' or `middle name', indicating a spurious match.

As was the case in the disambiguation of assignee names, in our final step we assign each name/geolocation pair a unique ID, ensuring that each matched name has the same ID.  From the 2,241,414 name/high-res geolocations found in the data, we produce 1,997,388 unique IDs after this round of disambiguation.  These identifiers are passed to the neighborhood search described in Sec. \ref{nearby.sec}.

\subsection{Nearby exact name matches}
\label{nearby.sec}

The disambiguations in Secs. \ref{assigneeExact.sec} and \ref{inventorExact.sec} provide a robust matching between similar names, but do so only at precise, high resolution geolocations.  Many addresses are low resolution, where e.g. the address ``Boston MA'' is very imprecise, and would not be used in the name disambiguation described in Secs. \ref{assigneeExact.sec} and \ref{inventorExact.sec}.  Any typographical errors in the address field that could also move precise addresses by even a few meters would prevent the linking of names as well (e.g. ``123 Main St.'' vs ``132 Main St.'').   In order to expand the coverage of the disambiguation to imprecise addresses, we perform a search for nearby geolocations that are found with the {\em{exact}} same name.  This is some sense the analog of the disambiguation described in Secs. \ref{assigneeExact.sec} and \ref{inventorExact.sec}, which used exact locations to search for similar names; now we use exact names to search for nearby locations.  

The names and geolocations (both high quality and low quality) for inventors and assignees are read into memory and processed as described in Sec. \ref{string.sec}.  For each unique name, only the highest quality geolocations are kept (e.g. for a name with two geolocations found using YQL, a location of quality 60 and another of quality 40, only the first would be kept).   To disambiguate the names:
\begin{enumerate}
\item{For each name occurring on any patent where YQL provided more than one geolocation of the same quality, we link all of those name/location pairs into the same identity.  This corrects any noise in the geolocation on an inventor's or assignee's address on a patent.}
\item{For each exact name, we determine the distance between all pairs of geolocations.  If that distance is less than 20km, link those name/locations.  }
\item{We assign all linked names to a unique identifier, and any alternate spellings of those names that were disambiguated following the steps in Secs. \ref{assigneeExact.sec} and \ref{inventorExact.sec} are {\em{also}} assigned to the same unique identifier.}
\item{Finally, we assign any unlinked names their own low-resolution unique identifier.}
\end{enumerate}
The output of this method is a list of 798,968 unique assignee IDs and 5,517,771 unique inventor IDs.  Many names are never found in conjunction with a high-precision geolocation, and misspellings or alternate spellings in these names will not be corrected using this algorithm.  For example, the patents US6495146 and US6028086  have assignee names ``Pfizer Incorporated'' and ``P Pfizer Inc'' respectively, both with the address ``New York, NY.''  These names will {\em{not}} be linked using this algorithm, due to the simultaneous imprecise naming {\em{and}} imprecise addresses.  Despite this limitation, we still have good coverage both in the number of disambiguated names and the number of disambiguated individuals and the number of patents they cover, as shown in Table \ref{disambigCount.tab}.  

\begin{table}[htdp]
\begin{center}
\begin{tabular}{|c|c|ccc|}
\hline
& number & EPO & PCT & USPTO\\
\hline
Disambiguated assignees with a  & 798,968 & 2,580,997 & 2,243,872 & 1,937,102 \\
geolocated address &  &  & &\\
\hline
Disambiguated inventors with a & 5,517,771 & 2,543,406 & 2,177,499 & 4,061,216\\
geolocated address & & & &\\
\hline
\hline
Disambiguated assignees with at least & 289,152 & 1,948,647 & 1,444,715 & 1,089,325 \\
one high-resolution geolocation & & (75\%) & (64\%) & (56\%)\\
\hline
Disambiguated inventors with at least & 2,080,080 & 1,957,999 & 1,398,891 & 1,873,018\\
one high-resolution geolocation & & (77\%) & (64\%) & (46\%)\\
\hline
\end{tabular}
\end{center}
\label{disambigCount.tab}
\caption{Disambiguated names of assignees and inventors that involved at least one high-resolution geolocation, meaning that the algorithm has potentially corrected for noise in the name as well as in the location.  The columns are the total number of disambiguated IDs, followed by the number of patents in each office linked to those IDs.  The percentages refer to the fraction of patents in each office {{with information on the address of the inventor or assignee}} that involved a name disambiguation using high precision addresses.  }
\end{table}%

We note that inventors or assignees without any provided address are not linked to any patents at all using this approach.  This is a more significant problem for patents in the USPTO than in the EPO or PCT, because in the US office there are $\sim$1.5M patents that have an assignee name but no assignee address provided.  

\subsection{Mobile Inventor Disambiguation}
\label{mobile.sec}

The approaches to disambiguation in Secs \ref{assigneeExact.sec}-\ref{nearby.sec} provide a disambiguation of institutions and inventors, but only within a $\sim$20km radius.  While assignees are generally expected to remain in a fixed and localized position (neglecting the possibility that the same company moves from one city to another), inventors are far more likely to move from one city to another.  Our algorithm would be incapable of linking the names of mobile inventors, since the distance between their geolocations would be well above our threshold of 20km.  In order to overcome this limitation, we must add one final step to the disambiguation algorithm that allows for inventor name matching over greater distances.  We do this by searching for exact name matches that are `similar' to one another in a manner other than distance.  In particular, we link a pair of disambiguated inventor IDs that share an {\em{exact}} name match if any of the patents held by those inventors
\begin{enumerate}
\item{share a disambiguated co-inventor}
\item{share a disambiguated co-assignee}
\item{are members of the same triadic family}
\item{have at least one citation from one to the other}
\end{enumerate}
A total of 2.7M pairs of potentially linked disambiguated IDs are observed in the data, where the same inventor name is found in distant locations.  Of those, 991,158 IDs satisfy at least one of the constraints for distant linking (i.e. the names share some other characteristic in common), and are linked using our algorithm.  This reduces 800,665 unique inventor IDs generated from the previous local disambiguation to 358,288 aggregated IDs that are used for the final inventor identifiers, a reduction of $\sim$8\% in the total number of inventor IDs.  Of those aggregated IDs, 188,726 ($\sim$53\% of the aggregated IDs) involve at least one inventor with a high-precision geolocation and 58,985 ($\sim$ 16\%) involve inventors that {\em{all}} have high resolution geolocations.   Inventors that can be linked to one high-precision geolocation are often focused on in the main text, and we emphasize that we do not require high resolutions in every city, but rather high resolutions in at least one city.  For example, two linked names that have the addresses ``Tokyo JP'' and ``123 Main Street, Boston MA USA'' {\em{would}} be labeled as a high-resolution inventor, since the address in Boston has the potential for being accurately disambiguated using our method.

\subsection{String Handling}
\label{string.sec}

When parsing the names, all latin characters are converted into lowercase letters.  All accented letters are converted into the character `x' (this has the affect of penalizing the removal of an accent, but not penalizing the change of an accent in a name).  All symbols (e.g. `-' or `.') are converted into spaces.  After this, words composed of a single character are deleted and all double-spaces, triple-spaces, etc. are converted to a space.

For assignees, a specific list of $\sim 30$ words are deleted from the names before passing through to disambiguation (for example, `inc', `corporation', `aktiengesellschaft', and `the').  The designation of words to completely delete before disambiguation were made manually, and altering this list may change the results to some extent.  

For inventor names, a few words are dropped from the names: ``dr'', ``de'', ``da'', ``di'' ``mc'', ``von'', ``der'', ``van'', and ``den''.  These short words tend to be non-informative within the last name of individuals, and cause problems with our algorithm if they are treated as a `first name.'

\end{document}